\documentclass[superscriptaddress,showpacs,twocolumn,prb]{revtex4}
\usepackage{amsmath}
\usepackage{amssymb}
\usepackage{graphicx}
\usepackage{dcolumn}
\usepackage{bm}

\begin{document}

\title{Anisotropic plasmons in a two-dimensional electron gas with
spin-orbit interaction}
\author{S. M. Badalyan}
\email{Samvel.Badalyan@physik.uni-regensburg.de}
\affiliation{Department of Radiophysics, Yerevan State University, 1 A. Manoukian St.,
Yerevan, 375025 Armenia}
\affiliation{Department of Physics, University of Regensburg, 93040 Regensburg, Germany}
\author{A. Matos-Abiague}
\affiliation{Department of Physics, University of Regensburg, 93040 Regensburg, Germany}
\author{G. Vignale}
\affiliation{Department of Physics and Astronomy, University of Missouri - Columbia,
Missouri 65211, USA}
\author{J. Fabian}
\affiliation{Department of Physics, University of Regensburg, 93040 Regensburg, Germany}
\date{\today}

\begin{abstract}
Spin-orbit coupling induced anisotropies of plasmon dynamics are
investigated in two-dimensional semiconductor structures. The interplay of
the linear Bychkov-Rashba and Dresselhaus spin-orbit interactions
drastically affects the plasmon spectrum: the dynamical structure factor
exhibits variations over several decades, prohibiting plasmon propagation in
specific directions. While this plasmon filtering makes the presence of
spin-orbit coupling in plasmon dynamics observable, it also offers a control
tool for plasmonic devices. Remarkably, if the strengths of the two
interactions are equal, not only the anisotropy, but all the traces of the
linear spin-orbit coupling in the collective response disappear.

\end{abstract}

\maketitle

Spin-orbit coupling in semiconductor heterostructures has received wide
attention recently -- it has been investigated as a source of new
fundamental spin physics as well as a control interaction in spintronics
applications \cite{zfds,fmesz}. Two spin-orbit terms are relevant in
zinc-blende systems exemplified by two-dimensional GaAs or InAs electron
gases: the Bychkov-Rashba~\cite{rashba} interaction (coupling constant $%
\alpha$), which is due to the structure inversion asymmetry, and the
Dresselhaus interaction~\cite{dresselhaus} (coupling constant $\beta$),
which is due to the bulk inversion asymmetry~\cite{fmesz}. Alone, these
interactions lead to an isotropic single-particle and plasmon spectrum.
Taken together, they imprint the underlying heterostructure anisotropy onto
the single and many-particle properties.
Most studies of the spin-orbit coupling effects have been on the
single-particle level. While the presence of spin-orbit coupling leads to
such notorious effects as spin relaxation \cite{fmesz} or spin Hall currents 
\cite{kato,wunderlich}, fascinating phenomena originate from the interplay
of the Bychkov-Rashba and Dresselhaus terms. The interplay often leads to
pronounced anisotropies 
\cite{golub,stano,matos,cheng,maytorena,bernevig2006,weber2007}, 
but this is not a rule \cite{trushin}. 

Recently several many-body effects important for spin properties of semiconductor nanosystems have been studied in 2DES~\cite{weber,smb,ulrich}. One of the key phenomena due to spin-orbit
interaction (SOI) in many-spin systems is the generation of the inter-chirality-subband electron-hole continuum. However, the dispersive and dissipative modifications, induced by individual (Bychkov-Rashba or Dresselhaus) SOI, are difficult to observe in experiment -- their effect is {\it isotropic} and proportional to the {\it small} SOI coupling~\cite{pletyukhov,kushwaha,saraga,wang,gumbs,xu,magarill,tse}. In real samples the interplay of different SOI mechanisms takes place and as we show here, it results in the striking anisotropy effect on the spectral properties of collective excitations in 2DES. This {\it qualitatively strong} effect can serve as a valuable tool to facilitate the observation and exploitation of usually weak SOI effects on many-body properties of 2DES.

An important outcome of our theory is the prediction of plasmon directional filtering: the interplay of the spin-orbit couplings leads to plasmon overdamping (blocking) in certain special directions of propagation and for certain magnitudes of the wave vector. This may be surprising at first sight, given that the spin-orbit effects on the plasmon dispersion and on the electron-hole excitation energies are in themselves quite small.  However even small energy shifts are sufficient, at these special wave vectors, to move the plasmon in or out of resonance with electron-hole excitations, thus producing a large effect on the plasmon damping. 
By scanning for
plasmons in different directions, this distinct absence of propagation in certain directions should be
experimentally verifiable, since the dynamical structure factor varies by
orders of magnitude as a function of the propagation angle. In addition to
making the spin-orbit presence experimentally visible, the anisotropy is
attractive for plasmonics designs as a substitute for surface patterning to
achieve directional plasmon propagation \cite{Maier06}. This prospect is
enforced by the possibility to control -- even turn on and off -- plasmon
propagation: both $\alpha $\ and $\beta $ can be tuned by external gates 
\cite{zfds} (see also \cite{ganichev,giglberger}) allowing for the
anisotropy to be tailored. In fact, the anisotropy vanishes (filtering
turned off) for $\alpha =\pm \beta $. More surprising, in this case the
(linear) spin-orbit couplings play no role in plasmon dynamics -- the
isotropic contributions by the individual spin-orbit terms cancel each other.

We calculate the effect of joint Bychkov-Rashba and Dresselhaus SOI on the propagation of plasmons in the (001) plane of a
zincblende semiconductor heterostructure.
We consider samples at low temperatures with high density 2DES where the kinetic energy of electrons dominates the Coulomb potential energy. In this regime it is legitimate to neglect the effect of exchange and correlations in treating plasmon excitations.
We use the random phase
approximation~\cite{GV05} and calculate the anisotropic Lindhard
polarization function for a given wave vector $\mathbf{q}$ and frequency $%
\omega$. The space in which the imaginary part of the Lindhard function
differs from zero is known as the electron-hole continuum (EHC)~\cite{GV05},
for it describes the spectrum of electron-hole excitations. The interplay of
the Bychkov-Rashba and Dresselhaus SOI leads to the appearance of several
sub-regions of the EHC separated by boundaries across which the imaginary
part of the dielectric function exhibits sharp variations. An interesting
effect arises when the frequency of a plasmon of a given $q$ but variable
propagation direction crosses these boundaries: The sudden rise in the
density of electron-hole excitations causes strong Landau damping, actually
overdamping the plasmons over a range of wave vector orientations. This
anisotropy of the plasmon spectrum should be observable through the
pronounced anisotropy of the dynamical structure factor, as shown below.

\begin{figure}[t]
\includegraphics[width=.95 \linewidth]{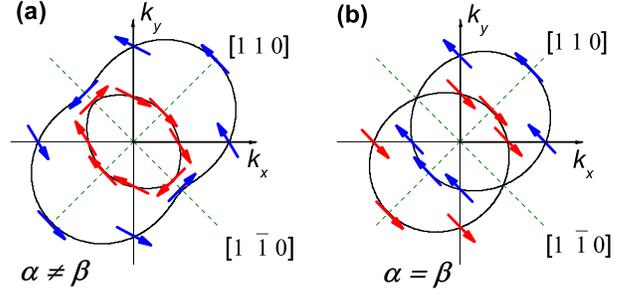}
\caption{Fermi contours in the momentum plane $(k_x,k_y)$ for a spin-orbit
interaction of the form given in Eq.~(\protect\ref{eq1}): (a) $\protect%
\alpha \neq \protect\beta$ and (b) $\protect\alpha = \protect\beta$. The
arrows indicate the spin direction. }
\label{fig1}
\end{figure}

Our spin-orbit interaction Hamiltonian is \cite{fmesz} 
\begin{equation}
H_{\text{SOI}}=\alpha \left( \hat{\sigma}_{x}k_{y}-\hat{\sigma}%
_{y}k_{x}\right) +\beta \left( \hat{\sigma}_{x}k_{x}-\hat{\sigma}%
_{y}k_{y}\right) ,  \label{eq1}
\end{equation}%
where $\hat{\sigma}_{x,y}$ are the Pauli matrices, $\vec{k}$ is the in-plane
electron momentum with magnitude $k$ and polar angle $\phi _{\mathbf{k}}$.
The eigenvectors of the Hamiltonian $H=H_{0}+H_{\text{SOI}}$ with $H_{0}={k}%
^{2}/2m^{\ast }$ ($m^{\ast }$ is the electron effective mass and $\hbar =1$)
are 
\begin{equation}
\Psi _{\mu }(\vec{r})=\frac{1}{\sqrt{2}}\left( 
\begin{array}{c}
ie^{-i\varphi } \\ 
\mu%
\end{array}%
\right) \frac{e^{i\vec{k}\vec{r}}}{\sqrt{A}},  \label{eq2}
\end{equation}%
corresponding to the single-particle spin-split branches of the electron
energy, 
\begin{equation}
E_{\mu }(\vec{k})=\frac{1}{2m^{\ast }}\left[ \left( k+\mu \ \xi (\rho
,\theta ,\phi _{\mathbf{k}})\right) ^{2}-\xi (\rho ,\theta ,\phi _{\mathbf{k}%
})^{2}\right] ~,  \label{eq3}
\end{equation}%
labeled by the chirality $\mu =\pm 1$; $A$ is the area of the 2DEG. The
phase of the spinor in Eq.~(\ref{eq2}) is $\varphi (\alpha ,\beta ,\phi _{%
\mathbf{k}})=\text{Arg}[\alpha e^{i\phi _{\mathbf{k}}}+i\beta e^{-i\phi _{%
\mathbf{k}}}]$ and the angle dependent Rashba-Dresselhaus momentum is%
\begin{equation}
\xi (\rho ,\theta ,\phi _{\mathbf{k}})=\rho \sqrt{1+\sin (2\theta )\sin
(2\phi _{\mathbf{k}})},  \label{eq4}
\end{equation}%
with amplitude $\rho =m^{\ast }\sqrt{\alpha ^{2}+\beta ^{2}}$. The angle
parameter $\theta $, defined as $\tan \theta =\beta /\alpha $, describes the
relative strength of the Bychkov-Rashba and Dresselhaus SOI. The Fermi
momenta of the subbands (\ref{eq3}) are also angle dependent: 
\begin{equation}
k_{F}^{\mu }(\rho ,\theta ,\phi _{\mathbf{k}})=\sqrt{2mE_{F}+\xi (\rho
,\theta ,\phi _{\mathbf{k}})^{2}}-\mu \ \xi (\rho ,\theta ,\phi _{\mathbf{k}%
})~,  \label{eq5}
\end{equation}%
where the total carrier density $n$ determines the Fermi energy, $%
E_{F}=\left( \pi n-\rho ^{2}\right) /m^{\ast }$. Figure~\ref{fig1}
illustrates the energy spectrum of the \textit{chirality} subbands and the
anisotropy of the Fermi contour (note that the Fermi energy can be negative).

The Lindhard polarization function \cite{GV05} in the presence of SOI is
defined as a sum over chirality indices $\Pi (\vec{q},\omega )=\sum_{\mu
,\nu =\pm 1}\Pi _{\mu \nu }(\vec{q},\omega )$, with%
\begin{eqnarray}
\Pi _{\mu \nu }(\vec{q},\omega ) &=&\int \frac{d\vec{k}}{\left( 2\pi \right)
^{2}}\frac{f[E_{\mu }(\vec{k})]-f[E_{\nu }(\vec{k}+\vec{q})]}{E_{\mu }(\vec{k%
})-E_{\nu }(\vec{k}+\vec{q})+\omega +i0}  \label{eq6} \\
&&\times \mathcal{F}_{\mu \nu }\left( \vec{k},\vec{k}+\vec{q}\right) ~, 
\notag
\end{eqnarray}%
where $f[E_{\mu }(\vec{k})]$ is the Fermi distribution function. The form
factors $\mathcal{F}_{\mu \nu }\left( \vec{k},\vec{k}+\vec{q}\right) $ are
given by%
\begin{equation}
\mathcal{F}_{\mu \nu }\left( \vec{k},\vec{k}+\vec{q}\right) =\frac{1}{2}%
\left[ 1+\mu \nu \cos \left( \Delta \varphi _{\mathbf{q}}\right) \right],
\label{eq7}
\end{equation}%
where we define $\Delta \varphi _{\mathbf{q}}=\varphi (\alpha ,\beta ,\phi _{%
\mathbf{k}})-\varphi (\alpha ,\beta ,\phi _{\mathbf{k}+\mathbf{q}})$. Notice
that in contrast to the case of pure Bychkov-Rashba or pure Dresselhaus SOI,
here the polarization function depends both on the magnitude, $q$, and
orientation, $\phi _{\mathbf{q}}$, of the wave vector $\vec{q}$. Making the
replacement $\vec{k}\rightarrow -\vec{k}-\vec{q}$ in the term of (\ref{eq7})
with $f[E_{\nu }(\vec{k}+\vec{q})]$ and regrouping, we can represent the
polarization function in the compact form $\Pi (\vec{q},\omega )=\sum_{\mu
,\nu ,\lambda =\pm 1}\Pi _{\mu ,\nu }^{\lambda }(\vec{q},\omega )$, where 
\begin{equation}
\Pi _{\mu ,\nu }^{\lambda }(\vec{q},\omega )=\int \frac{d\vec{k}}{\left(
2\pi \right) ^{2}}\frac{f[E_{\mu }(\vec{k})]\mathcal{F}_{\mu \nu }\left( 
\vec{k},\vec{k}+\vec{q}\right) }{E_{\mu }(\vec{k})-E_{\nu }(\vec{k}+\vec{q}%
)+\lambda \left( \omega +i0\right) }~.  \label{eq8}
\end{equation}

Exploiting further the symmetry of the problem, in the limit of zero
temperature we reduce the polarization function to the following expression 
\begin{equation}
\Pi (\vec{q},\omega )=\frac{g}{4\pi }\sum\limits_{\mu ,\lambda
}\int\limits_{0}^{2\pi }d\phi _{\mathbf{k}}\int\limits_{0}^{v_{F,\mu }}dv%
\frac{v(e_{\mu ,\lambda }-d_{\mu }v)}{a_{\mu }(v-v_{\mu ,\lambda
}^{+})(v-v_{\mu ,\lambda }^{-})}~.  \label{eq9}
\end{equation}%
Here we have defined the dimensionless Fermi wave vector $v_{F,\mu }=\sqrt{%
1-r^{2}+\overline{\xi }_{\mathbf{k}}^{2}}-\mu \overline{\xi }_{\mathbf{k}}$,
and the functions $v_{\mu ,\lambda }^{\pm }=\left( -b_{\mu ,\lambda }\pm 
\sqrt{{b_{\mu ,\lambda }^{2}}-4a_{\mu }c_{\lambda }}\right) /2a_{\mu }$,
with 
\begin{eqnarray}
a_{\mu } &\equiv &x\cos \left( \phi _{\mathbf{k}}-\phi _{\mathbf{q}}\right) %
\left[ x\cos \left( \phi _{\mathbf{k}}-\phi _{\mathbf{q}}\right) -\mu 
\overline{\xi }_{\mathbf{k}}\right] ~;  \label{eq10} \\
b_{\mu ,\lambda } &\equiv &-x\left[ \left( r^{2}+2(\lambda y-x^{2})\right)
\cos \left( \phi _{\mathbf{k}}-\phi _{\mathbf{q}}\right) \right.
\label{eq11} \\
&+&\left. r^{2}\sin \left( 2\theta \right) \sin \left( \phi _{\mathbf{k}%
}+\phi _{\mathbf{q}}\right) \right] +\mu (\lambda y-x^{2})\overline{\xi }_{%
\mathbf{k}}~;  \notag \\
c_{\lambda } &\equiv &\left( \lambda y-x^{2}\right) ^{2}-x^{2}\overline{\xi }%
_{\mathbf{q}}{}^{2}~;  \label{eq12} \\
d_{\mu } &\equiv &x\cos \left( \phi _{\mathbf{k}}-\phi _{\mathbf{q}}\right)
-\mu \overline{\xi }_{\mathbf{k}}~;  \label{eq13} \\
e_{\mu ,\lambda } &\equiv &\lambda y-x^{2}  \label{eq14} \\
&+&\frac{\mu r^{2}x}{\overline{\xi }_{\mathbf{k}}}\left[ \cos \left( \phi _{%
\mathbf{k}}-\phi _{\mathbf{q}}\right) +\sin \left( \phi _{\mathbf{k}}+\phi _{%
\mathbf{q}}\right) \sin (2\theta )\right] ~.  \notag
\end{eqnarray}%
Here $g=m^{\ast }/2\pi $ is the density of states at the Fermi level and we
have introduced the dimensionless quantities $x=q/2k_{F}$, $y=\omega
/4\varepsilon _{F}$, $v=k/k_{F}$, $r=\rho /k_{F}$, and $\overline{\xi }_{%
\mathbf{k}}=\xi (\rho ,\theta ,\phi _{\mathbf{k}})/k_{F}$ with $\varepsilon
_{F}=k_{F}^{2}/2m^{\ast }$ and $k_{F}=\sqrt{2m^{\ast }E_{F}+\rho ^{2}}$~.
The integration over $v$ can be done analytically, yielding 
\begin{align}
& \Pi (\vec{q},\omega )=-\frac{g}{4\pi }\sum_{\mu ,\lambda }\int_{0}^{2\pi
}d\phi _{\mathbf{k}}\left\{ \frac{dv_{F}}{a}+\frac{1}{a\left(
v^{+}-v^{-}\right) }\right.  \label{eq15} \\
& \times \left. \left[ v^{+}(e-dv^{+})\ln \frac{v^{+}}{v^{+}-v_{F}}%
-v^{-}(e-dv^{-})\ln \frac{v^{-}}{v^{-}-v_{F}}\right] \right\} ~.  \notag
\end{align}

\begin{figure}[t]
\includegraphics[width=.53\linewidth]{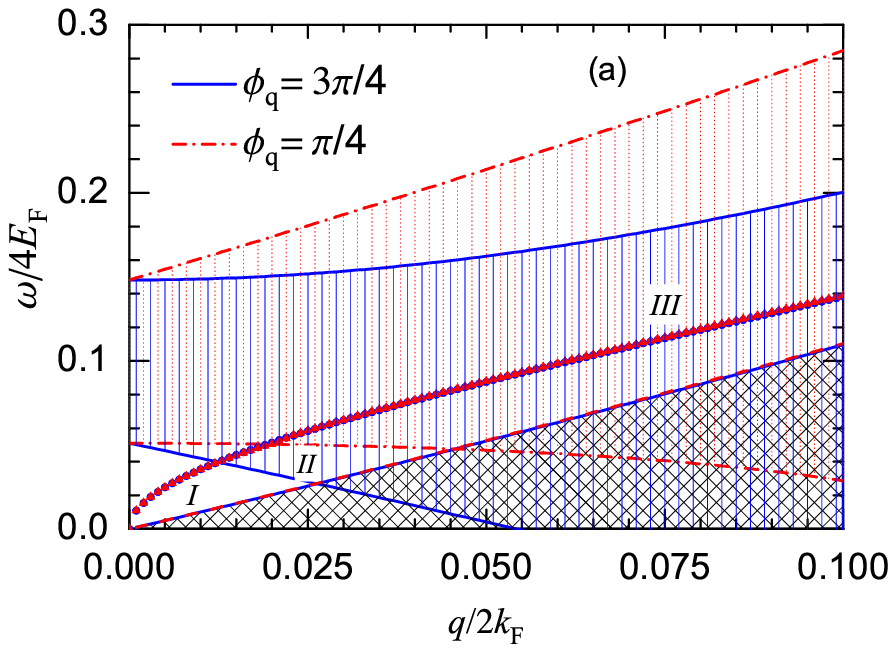} \includegraphics[width=.45%
\linewidth]{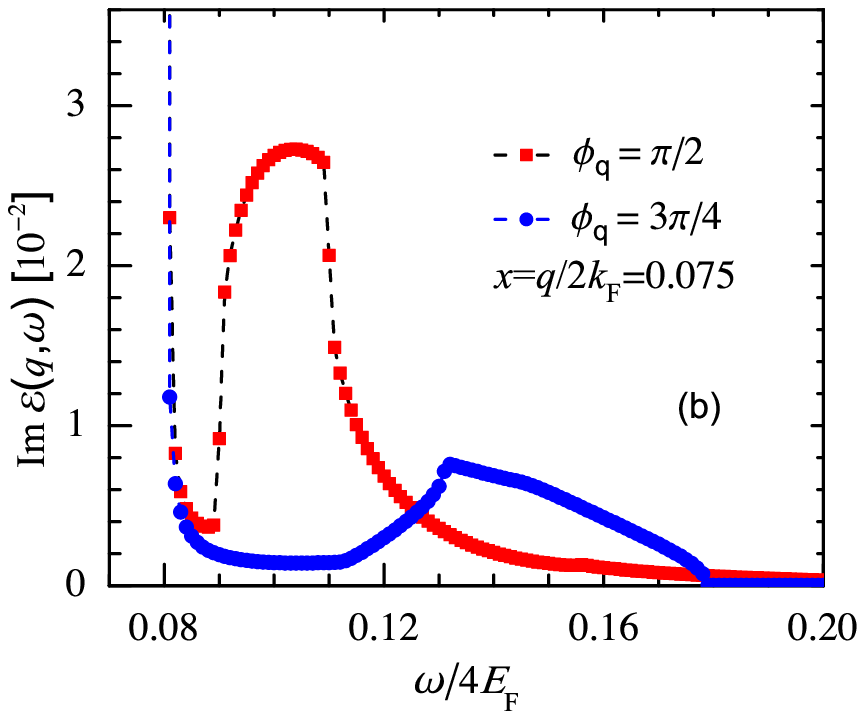}
\caption{(a) The intra- and inter-chirality EHC in the $\protect\omega -q$
plane for two different momentum orientations, $\protect\phi _{\mathbf{q}}=%
\protect\pi /4$ and $3\protect\pi /4$. The symbols show the plasmon
dispersions (see text). (b) The imaginary part of the dielectric function vs
energy for the fixed momentum magnitude $q=0.15k_{F}$ and for $\protect\phi %
_{\mathbf{q}}=\protect\pi /2$ and $3\protect\pi /4$, shown as square and
round symbols, respectively. }
\label{fig2}
\end{figure}

The integration over $\phi _{\mathbf{k}}$ is performed numerically. The
derived formula (\ref{eq15}) is exact. In the limit of $\theta =0$ (pure
Bychkov-Rashba SOI) we recover the previous results by Pletyukhov and
Gritsev~\cite{pletyukhov}, and in the limit of $r=0$ (no SOI) we recover the
classic result by Stern~\cite{stern}.

For the actual calculations we use materials parameters suitable for InAs
quantum wells with realistic values of the SOI parameters, $r=0.1$ and $%
\theta =\pi /8$, corresponding to the ratio of SOI strengths $\alpha /\beta
\approx 2.4$ from~\cite{giglberger}. We take the electron density $%
n=2.55\cdot 10^{11}$~cm$^{-2}$ ($E_{F}\approx 302$ K) and the effective
transverse width of the quantum well $d=15$~nm.

Figure~\ref{fig2}a shows the EHC regions and the plasmon dispersions for
different values of the angle $\phi _{\mathbf{q}}$. The anisotropy of the 
\emph{intra}-chirality EHC (the dense-hatched region) and of the plasmon
dispersions is a small effect and hardly seen on the scale of figure.
Meantime, the \emph{inter}-chirality EHC is strongly anisotropic (in the
long wavelength limit the anisotropy vanishes). Figure~\ref{fig2}b shows the
imaginary part of the dielectric function vs energy for the fixed momentum
magnitude and different orientations. As seen, not only the boundaries of
EHC but also the dissipation properties \emph{within EHC} are strongly
anisotropic. In the region near the plasmon energy, $\omega /4E_{F}\approx
0.1$ for $q=0.15k_{F}$, the imaginary part for $\phi _{\mathbf{q}}=3\pi /4$
is strongly suppressed with respect to its value for $\phi _{\mathbf{q}}=\pi
/2$.

To calculate the plasmon dispersion we solve for zeros of the real part of
the RPA dielectric function, $\varepsilon (\vec{q},\omega )=1-v(q)\Pi (\vec{q%
},\omega )$ where $v(q)=2\pi e^{2}/(\kappa _{0}q)F(q d)$ is the bare Coulomb
interaction with $\kappa _{0}=14.55$ the static dielectric constant of InAs. For the form factor $F(q d)$ we use the formula (12) from Ref.~\onlinecite{bkvs2007}, which takes into account the transverse width $d$ of the quantum well but not its asymmetric shape. The form factor has a strong effect on the Coulomb interaction strength~\cite{smb} --  it goes as  $1-(1/3-5/4\pi^2)q d$ in the long wavelength limit $q d\rightarrow 0$,  and as $3/(4\pi^2 q d)$, in the opposite limit $q d\rightarrow \infty$.

\begin{figure}[t]
\includegraphics[width=.75\linewidth]{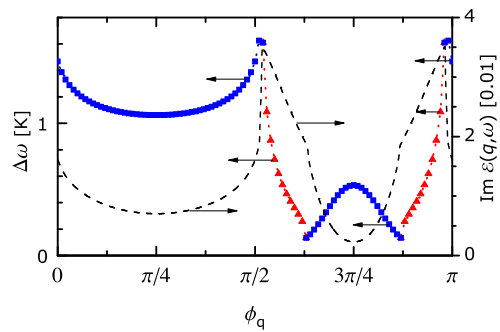}
\caption{The SOI induced energy dispersion of the plasmon vs its propagation
direction for $q=0.15k_{F}$ (the left axis, square symbols). The parts of
the spectrum which \textit{do not} represent plasmon excitations (see text)
are shown as triangle symbols. The dashed line plots the imaginary part of
the dielectric function (the right axis) for $q=0.15k_{F}$ and $\Delta 
\protect\omega \equiv \protect\omega -\protect\omega _{0}=1.2$ K where $%
\protect\omega _{0}\approx 0.45E_{F}$.}
\label{fig3}
\end{figure}

There are three distinct regions of EHC and the plasmon disperions, as seen
in Fig.~\ref{fig2}a. In region \textit{I}, which corresponds to small $q$,
the areas of inter- and intra-chirality subband transitions are well
separated. The plasmon energy is located within the gap between these EHC
regions: these plasmons are not dampled. The plasmons here exhibit only a
SOI induced dispersion as a function of its propagation orientation. At such
small $q$, however, the anisotropy is not significant and eventually
vanishes in the long wavelength limit.

At larger values of $q$, in the region \textit{III} in Fig.~\ref{fig2}a, the
plasmon dispersion enters EHC, triggering the phenomenon known as Landau
damping, i.e. decay into electron-hole pairs. In this regime the EHC is made
up of several overlapping sub-regions (associated with the discrete quantum
indices $\mu $ and $\lambda $), separated by sharp boundaries. The imaginary
part of the dielectric function (proportional to the spectral density of
electron-hole pairs) exhibits sharp variations across these boundaries,
resulting in unexpectedly strong angular dependence of plasmon damping. In
Fig.~\ref{fig3} we follow the evolution of the plasmon frequency as a
function of the angle $\phi _{q}$ from $0$ to $\pi $ for $q=0.15k_{F}.$ A
sharp boundary between two sub-regions of the EHC is crossed at $\phi
_{q}\simeq \pi /2$. Entering the new region, the plasmon becomes overdamped,
concurrent with the sharp rise of Im~$\varepsilon (\vec{q},\omega )$, which
we plot in the same figure on the right axis. Figure~\ref{fig3} shows that
there are two ranges of directions $\pi /2\overset{<}{\sim }\phi _{q}\overset%
{<}{\sim }5\pi /8$ and $7\pi /8\overset{<}{\sim }\phi _{q}\overset{<}{\sim }%
\pi $ in which the plasmon cannot propagate due to excessive Landau damping.
On the other hand, the plasmon is well defined around the angles $\phi _{%
\mathbf{q}}=\pi /4$ and $3\pi /4$ where the imaginary part of $\varepsilon (%
\vec{q},\omega )$ is small. These are the principal directions of the
underlying structural $C_{2v}$ symmetry.

Finally, in the intermediate \textit{II} region in Fig.~\ref{fig2}, the
intra- and inter-chirality subbands either overlap or not so that the
plasmon is being either damped or not depending on its propagation direction.

\begin{figure}[t]
\includegraphics[width=.75\linewidth]{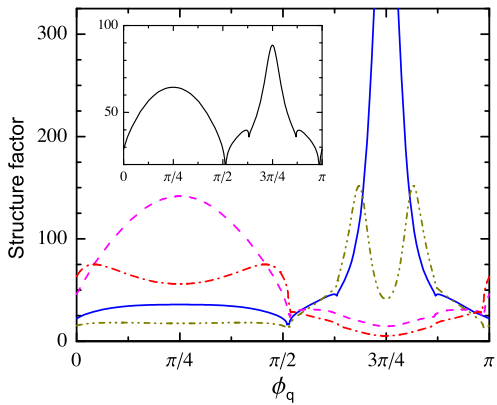}
\caption{The dynamical structure factor vs $\protect\phi_{\mathbf{q}}$ for $%
q = 0.15 k_F$. The solid and dashed lines correspond, respectively, to the
local maximum at $\Delta\protect\omega \approx 0.5$~K and minimum at $\Delta%
\protect\omega\approx 1.1$~K of the plasmon spectrum (cf. Fig.~\protect\ref%
{fig3}) and show the single-peak behavior of $S(\vec{q},\protect\omega )$.
The dashed-dot-dot and dashed-dot lines illustrate the splitting of the
structure factor peaks for $\Delta\protect\omega \approx 0.25$ and $1.5$ K.
The inset shows the asymmetric double-peak structure of the structure factor
for $\Delta\protect\omega \approx 0.7$ K. }
\label{fig4}
\end{figure}

In Fig.~~\ref{fig4} we plot the dynamical structure factor $S(\vec{q},\omega
)=-\text{Im}\left[ 1/\varepsilon (\vec{q},\omega )\right] $ as a function of 
$\phi _{\mathbf{q}}$ for $\omega $ corresponding to the local maximum and
minimum of the plasmon energy dispersion. In both cases $S(\vec{q},\omega )$
shows a dominant peak as a function of $\phi _{q}$. As expected, the peak
occurs at $\phi _{\mathbf{q}}=\pi /4$ for $\omega =\omega _{min}(q)$ (dashed
line) and at $\phi _{q}=3\pi /4$ for $\omega =\omega _{max}(q)$ (solid
line). These peaks represent lightly Landau damped plasmons (the plasmon at $%
\phi _{q}=3\pi /4$ is less damped than the one at $\phi _{q}=\pi /4$ and
therefore produces a stronger peak). The preferential role of these two
directions comes from the $C_{2v}$ symmetry of the problem, clearly seen
from the plot of the Fermi surface in Fig.~1. Notice that for a given $%
\omega $ there are two additional angles, at which Re~$\varepsilon (\vec{q}%
,\omega )$ shows zeros. The structure factor, however, does not exhibit
peaks at these angles since the large density of electron-hole pairs
(reflected in the large value of Im~$\varepsilon (\vec{q},\omega )$)
overdamps the plasmons in these directions. These \textquotedblleft
overdamped plasmons" are represented by the triangle symboles in Fig.~\ref%
{fig3}.

In the range $\omega_{max}(q)<\omega<\omega_{min}(q)$ between the extrema of
the plasmon spectrum, the height and the width of the peaks of $S(\vec{q}%
,\omega )$ vs $\phi_q$ show a smooth evolution: with increasing $\omega$ one
peak diminishes, the other grows, and vice versa. Thus, in this intermediate
region the structure factor has two peaks, located at $\phi _{\mathbf{q}%
}=\pi /4$ and $3\pi /4$, which constitute an asymmetric doublet, shown in
the inset of Fig.~\ref{fig4}. In the energy regions above the minimum or
below the maximum of the plasmon spectrum at given $q$ (i.e., for $\omega
>\omega _{min}(q)$ or $\omega <\omega _{max}(q)$), Re~$\varepsilon (\vec{q}%
,\omega )$ vs $\phi_{\mathbf{q}} $ shows two zeros around $\pi /4$ or $3\pi
/4$ so that each peak of $S(\vec{q},\omega )$ splits into two peaks located
symmetrically above and below the angle $\phi_{\mathbf{q}}=\pi /4$ (the
dash-dot line) or $3\pi /4$ (the dash-dot-dot line).

In the case of $\alpha =\pm \beta $ (see Fig.~\ref{fig1}b) the linear
spin-orbit couplings do not affect the plasmon spectrum: the plasmon damping
vanishes and the structure factor is a delta-function for all momentum
orientations.  For this special case there is a global spin quantization
axis -- one of the principal $C_{2v}$ axes -- so that the electron gas is
split into two uncoupled spin components, whose circular Fermi contours are
shifted from the origin in opposite directions. Each component gives an
isotropic collective response, as guaranteed by Galilean invariance. Cubic
spin-orbit terms, which spoil this effect, are typically much weaker in
quantum wells.

In conclusion, we have shown that plasmon dynamics (spectrum and damping) is
strongly anisotropic in realistic zinc-blende quantum wells, due to the
interplay of two different forms of spin-orbit interaction. Experimental
observation of this anisotropy would be of fundamental interest and could
open the way to new techniques for controlled directional plasmon filtering
potentially useful both for spintronic and plasmonic devices.

This work is supported by the Volkswagen Foundation, the SFB Grant 689,
NSF Grant No. DMR-0705460, and ANSEF grant. 
We thank T. Reinecke, C. Schuller, T. Korn, and S. Abedinpour for useful 
discussions.

\end{document}